\documentclass[conference]{IEEEtran}
\IEEEoverridecommandlockouts

\newlength\tindent
\renewcommand{\indent}{\hspace*{\tindent}}
\usepackage{amsmath,amssymb,amsfonts}
\usepackage{graphicx}
\usepackage{stmaryrd}
\usepackage{textcomp}
\usepackage{xcolor}
\usepackage{comment}
\def\BibTeX{{\rm B\kern-.05em{\sc i\kern-.025em b}\kern-.08em
    T\kern-.1667em\lower.7ex\hbox{E}\kern-.125emX}}

\usepackage[T1]{fontenc}		
\usepackage[utf8]{inputenc}		
\usepackage[english]{babel}		
\usepackage{geometry}			
\usepackage{setspace}			
\usepackage{fancyhdr}			

\usepackage{afterpage} 			
\usepackage{hyperref}			
\usepackage{url}
\usepackage{color}			
\usepackage{enumerate}			
\usepackage{enumitem}			
\usepackage{pstool}			
\usepackage{psfrag}
\usepackage{subcaption}
\usepackage{tabularx}			
\usepackage[bottom]{footmisc}		
\usepackage{booktabs}			
\usepackage{longtable}			

\usepackage{cellspace} 
\usepackage{caption}			
\usepackage{float}			
\usepackage[countmax]{subfloat}			
\usepackage{rotating}			
\usepackage{rotfloat}			
\usepackage{amsmath} 			
\usepackage{amssymb}			
\usepackage{cancel}                     
\usepackage{mleftright}
\usepackage{listings}                   
\usepackage[swapnames,norules,nouppercase]{frontespizio}
\usepackage[intoc]{nomencl}
\usepackage[acronym,toc,nomain,nopostdot,nonumberlist]{glossaries}
\usepackage[square, numbers, comma, sort&compress]{natbib}

\usepackage{verbatim}                   
\usepackage{wrapfig}                 
\usepackage{blindtext}

\usepackage{calligra}

\usepackage{microtype}
\usepackage{graphicx}
\usepackage{tcolorbox} 
\usepackage[normalem]{ulem}
\usepackage{mathtools}
\usepackage{amsfonts}
\usepackage{bm}
\usepackage{amsthm}
\usepackage[noabbrev,capitalise,nameinlink]{cleveref}
\hypersetup{colorlinks={true},linkcolor={darkred},citecolor=darkblue}

\usepackage{algorithm}
\usepackage[noend]{algpseudocode}
\newcommand{\algalign}[2]

\newlength{\bibitemsep}
\newlength{\bibparskip}
\let\oldthebibliography\thebibliography
\renewcommand\thebibliography[1]{%
  \oldthebibliography{#1}%
  \setlength{\parskip}{\bibitemsep}%
  \setlength{\itemsep}{\bibparskip}%
}

\usepackage{dsfont}

\usepackage{stmaryrd} 
\usepackage{tikz} 
\usepackage{tikz-cd} 
\usetikzlibrary{arrows.meta, calc, positioning, decorations.markings}
\usepackage{multirow}
\usepackage{changepage}
\usepackage{threeparttable}
\usepackage{makecell}
\usepackage{xspace}
\usepackage[colorinlistoftodos]{todonotes}
\usepackage{adjustbox}

\usepackage{comment}

\fancypagestyle{plain}{
                        \fancyhf{}                              
                        \fancyfoot[R]{\thepage}                 
                        }                                       
\definecolor{sapienza}{RGB}{130,36,51} 
\definecolor{cust1}{RGB}{85,85,85}
\definecolor{cust2}{RGB}{212,212,212}
\makenomenclature 
\makeglossaries 

\definecolor{bluegray}{rgb}{0.4, 0.6, 0.8}
\definecolor{electriclime}{rgb}{0.8, 1.0, 0.0}
\definecolor{malachite}{rgb}{0.04, 0.85, 0.32}
\definecolor{darkred}{rgb}{0.55, 0.0, 0.0}
\definecolor{darkblue}{rgb}{0.0, 0.0, 0.55}
\definecolor{darkgreen}{rgb}{0.0, 0.2, 0.13}
\definecolor{darkorchid}{rgb}{0.6, 0.2, 0.8}
\definecolor{simplexcolor}{RGB}{151,204,200}

\newcommand{\V}{\mathsf{V}}
\newcommand{\E}{\mathsf{E}}

\newcommand{\R}{\mathbb{R}}

\newcommand{\G}{\mathsf{G}}
\newcommand{\GNN}{\mathsf{GNN}}     

\theoremstyle{plain}

\theoremstyle{definition}

\theoremstyle{remark}

\definecolor{gold}{RGB}{221, 196, 65}
\definecolor{silver}{RGB}{215, 215, 215}
\definecolor{bronze}{RGB}{205, 127, 50}

\definecolor{lightgray}{gray}{0.95}
\definecolor{Gray}{gray}{0.85}
\definecolor{LightCyan}{rgb}{0.88,1,1}
\definecolor{LightPink}{HTML}{FCE1EF}
\definecolor{LightGreen}{HTML}{EEF7E1}
\newcolumntype{A}{>{\columncolor{white}}c}
\newcolumntype{B}{>{\columncolor{LightGreen}}c}
\newcolumntype{C}{>{\columncolor{LightPink}}c}

\definecolor{purpleheart}{rgb}{0.41, 0.21, 0.61}
\definecolor{mblue}{rgb}{0.2118, 0.3608, 0.5765}
\definecolor{mgreen}{rgb}{0.2314, 0.5765, 0.3961}
\definecolor{mred}{rgb}{0.5765, 0.2314, 0.3216}
\definecolor{mpurp}{rgb}{0.4824, 0.2314, 0.5765}
\definecolor{amber}{rgb}{1.0, 0.75, 0.0}

\definecolor{dark2green}{rgb}{0.1, 0.65, 0.3}
\definecolor{dark2orange}{rgb}{0.9, 0.4, 0.}
\definecolor{dark2purple}{rgb}{0.4, 0.4, 0.8}

\definecolor{lightgreen}{RGB}{168,207,147}
\definecolor{lightblue}{RGB}{34,118,180}
\definecolor{lightyellow}{RGB}{255,226,149}

\begin{document}

\title{Towards Explainable Graph Neural Networks for Neurological Evaluation on EEG Signals}

\author{
\IEEEauthorblockN{Andrea Protani\IEEEauthorrefmark{1}\IEEEauthorrefmark{2}, 
Lorenzo Giusti\IEEEauthorrefmark{1}, 
Chiara Iacovelli\IEEEauthorrefmark{4}, 
Albert Sund Aillet\IEEEauthorrefmark{1}, \\
Diogo Reis Santos\IEEEauthorrefmark{1}, 
Giuseppe Reale\IEEEauthorrefmark{4}, 
Aurelia Zauli\IEEEauthorrefmark{4}, 
Marco Moci\IEEEauthorrefmark{4}, \\
Marta Garbuglia\IEEEauthorrefmark{4}, 
Pierpaolo Brutti\IEEEauthorrefmark{2}, 
Pietro Caliandro\IEEEauthorrefmark{4}, 
Luigi Serio\IEEEauthorrefmark{1}} \\
\IEEEauthorblockA{\IEEEauthorrefmark{1}CERN, Geneva, Switzerland \\
\{andrea.protani, lorenzo.giusti, albert.sund.aillet, diogo.reis.santos, Luigi.Serio\}@cern.ch}
\IEEEauthorblockA{\IEEEauthorrefmark{2}Sapienza University of Rome, Rome, Italy \\
pierpaolo.brutti@uniroma1.it}
\IEEEauthorblockA{\IEEEauthorrefmark{4}Fondazione Policlinico Universitario Agostino Gemelli IRCCS, Rome, Italy \\
\{chiara.iacovelli, giuseppe.reale, pietro.caliandro\}@policlinicogemelli.it \\
\{aurelia.zauli, marco.moci, marta.garbuglia\}@guest.policlinicogemelli.it}
}

\maketitle

\begin{abstract}
After an acute stroke, accurately estimating stroke severity is crucial for healthcare professionals to effectively manage patient's treatment. Graph theory methods have shown that brain connectivity undergoes frequency-dependent reorganization post-stroke, adapting to new conditions. Traditional methods often rely on handcrafted features that may not capture the complexities of clinical phenomena. In this study, we propose a novel approach using Graph Neural Networks (GNNs) to predict stroke severity, as measured by the NIH Stroke Scale (NIHSS). We analyzed electroencephalography (EEG) recordings from 71 patients at the time of hospitalization. For each patient, we generated five graphs weighted by Lagged Linear Coherence (LLC) between signals from distinct Brodmann Areas, covering $\delta$ (2-4 Hz), $\theta$ (4-8 Hz), $\alpha_1$ (8-10.5 Hz), $\alpha_2$ (10.5-13 Hz), and $\beta_1$ (13-20 Hz) frequency bands. To emphasize key neurological connections and maintain sparsity, we applied a sparsification process based on structural and functional brain network properties. We then trained a graph attention model to predict the NIHSS. By examining its attention coefficients, our model reveals insights into brain reconfiguration, providing clinicians with a valuable tool for diagnosis, personalized treatment, and early intervention in neurorehabilitation.
\end{abstract}

\begin{IEEEkeywords}
Clinical Neuroscience, Stroke, Graph Neural Networks, Explainable AI, NIHSS.
\end{IEEEkeywords}

\section{Introduction}

\begin{figure*}[h]
    \centering
    \vspace{-5mm}
    \includegraphics[width=0.75\linewidth]{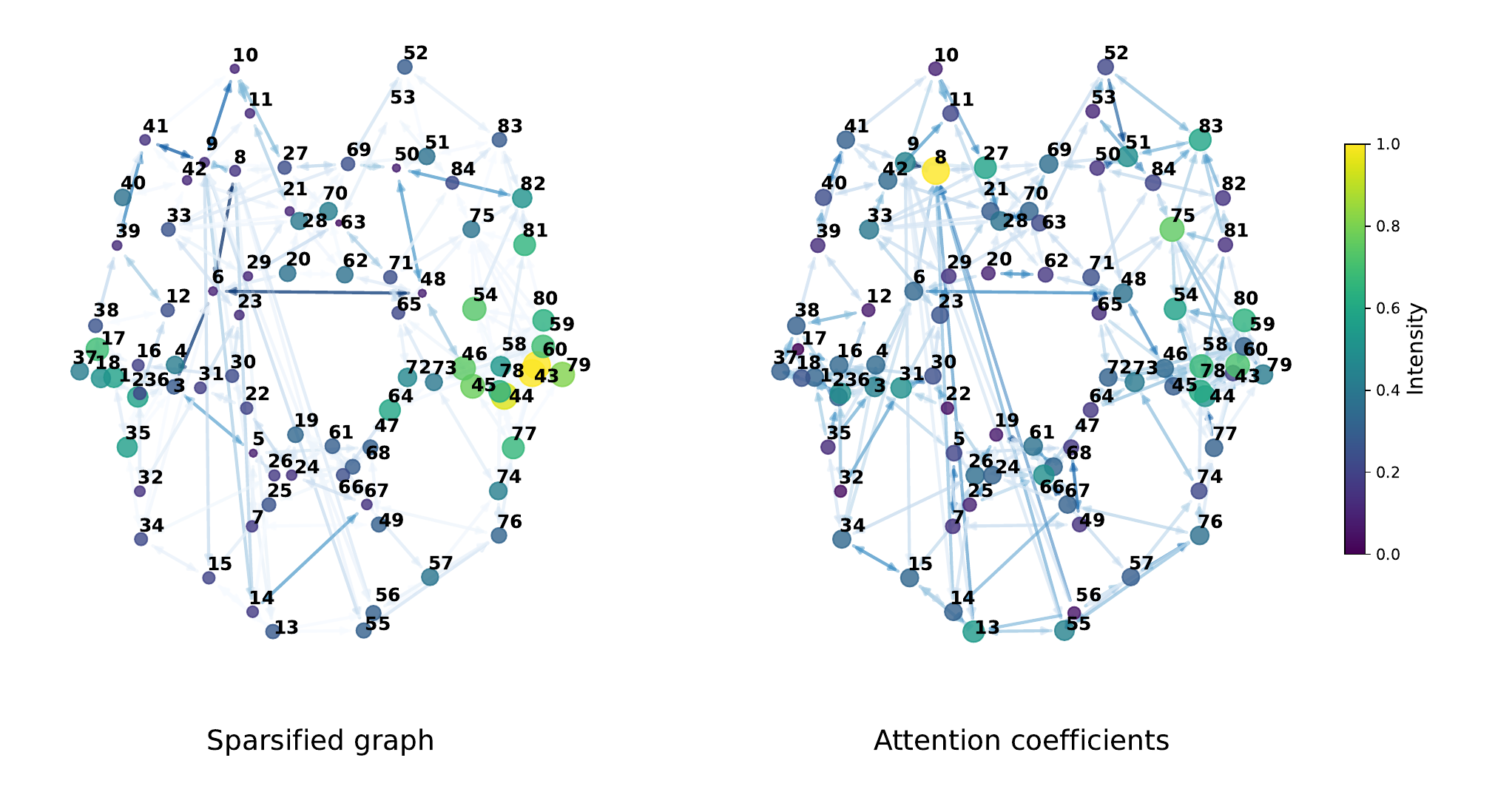}
    \caption{Patient A, Stroke side, right - NIHSS, 19: (Left) Brain graph analysis based on small-world metrics, specifically using the weighted clustering coefficient for the nodes and edge betweenness for the edges. (Right) Nodes are visualized using weighted in-degree centrality calculated through attention coefficients, while the edges represent the actual values of the attention coefficients. These are the combined graph built from $\G_{\alpha_1}$, $\G_{\alpha_2}$, $\G_{\beta_1}$. For each pair of Brodmann areas, the edge with the highest value was selected: $e_{ij} = \max(e_{ij}^{\alpha_1}, e_{ij}^{\alpha_2}, e_{ij}^{\beta_1})$.}
     \label{fig:g_theory_vs_attn}
     \vspace{-5mm}
\end{figure*}

Clinical scientists have observed that after cerebral ischemia, the brain's functional connectivity is restructured both locally and remotely from the stroke site~\cite{wang2010dynamic, crofts2011network}. Using tools from graph theory principles~\cite{bassett2017network} allows modeling anatomical brain regions as nodes and their functional connectivity as edges to provide quantitative analysis of structural and functional changes post-acute ischemic stroke~\cite{caliandro2017small, guggisberg2019brain}. This approach has unveiled critical insights into how stroke affects brain network properties such as small-worldness, efficiency, and modularity, essential for understanding the brain's adaptive mechanisms during recovery~\cite{vecchio2019cortical, vecchio2019acute, vecchio2023prognostic}. Meanwhile, recent advancements in graph representation learning have enhanced our ability to analyze and identify patterns in graph-structured data~\cite{gilmer2017neural, kipf2017graph, velivckovic2018graph}. In this paper, we merge clinical neuroscience with GNNs to propose an explainable model for objectively quantifying stroke impairment using the NIH Stroke Scale (NIHSS). Our approach employs graph attention models to predict the clinical severity score, according to the NIHSS, and identify key brain regions involved in the prediction, enhancing interpretability for clinical application. As illustrated in~\Cref{fig:g_theory_vs_attn}, our model's explanations align with clinical neuroscience literature, highlighting significant neurological connections consistent with known stroke recovery mechanisms. To our knowledge, this is the first application of Graph Neural Networks on EEG connectivity data of acute ischemic stroke patients during hospitalization  in stroke units, opening the way for more accurate diagnoses and therapeutic strategies in clinical neuroscience.\\

\begin{figure*}[t]
    \centering
    \includegraphics[width=0.9\linewidth]{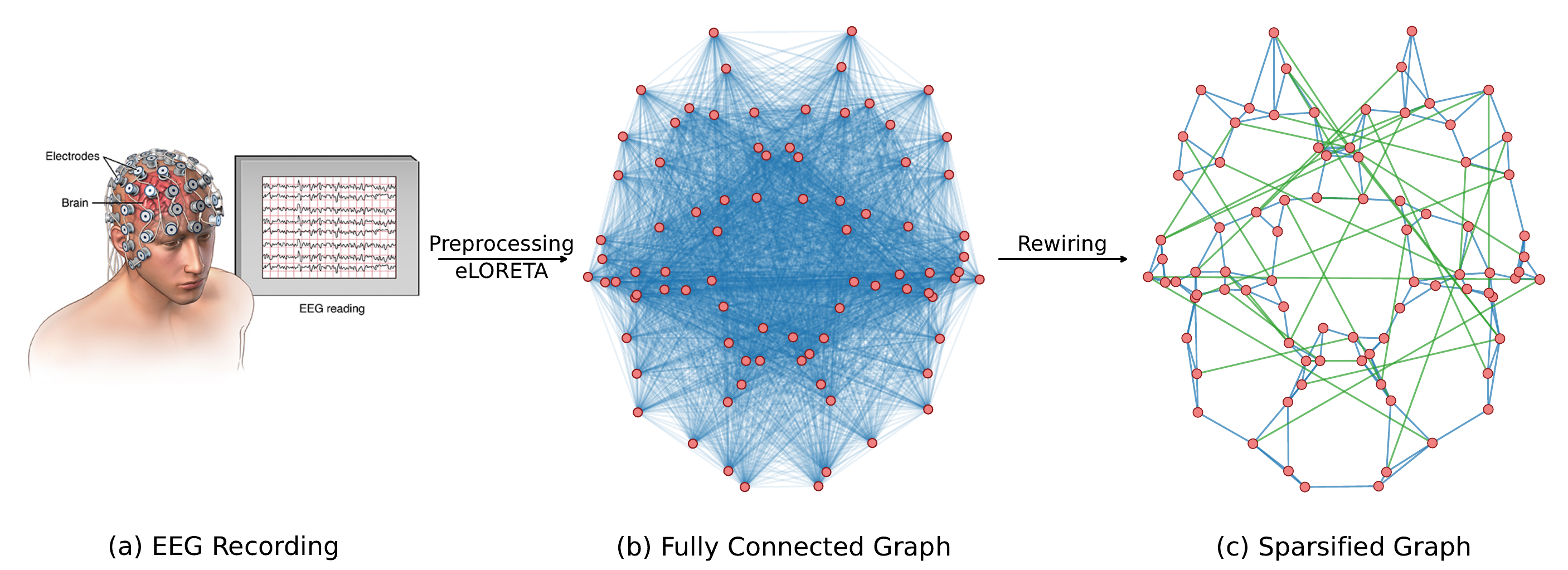}
     \caption{Full pipeline from EEG data collection to model input: 
  (a) EEG recording setup, illustrating the placement of electrodes on the scalp for data collection~\cite{Shen_2020}. 
  (b) The initial fully connected graph obtained after the EEG's preprocessing and the application of eLORETA, representing all possible connections between 84 Brodmann areas in the brain, constructed from the EEG data. 
  (c) Sparsified graph after the rewiring process, serving as input to the model. Blue edges represent structural connections based on spatial proximity, while green edges are functional connections derived from the top 1\% of LLC values.}
     \label{fig:brain_network_plot}
\end{figure*}

\section{Materials \& Methods}

\textbf{Graph Neural Networks:} A graph $\G = (\V, \E)$ consists of a set $\V$ of nodes (vertices) representing entities and a set $\E$ of edges (links) encoding the relationships between them. For any two nodes $u, v \in \V$, an edge exists if $(u,v) \in \E$~\citep{bondy2008graph}. The connectivity of graph $\G$ is described by the adjacency matrix $\mathbf{A}\in\mathbb{R}^{n\times n}$, where $n$ is the number of nodes. We assume $\G$ is undirected and connected, with features $\{{\mathbf{x}}_{v}\}_{v\in V} \in \R^{d}$. GNNs are functions of the form ${\GNN}{\mathbf{w}}:(\G, \{{\mathbf{x}}_{v}\}_{v\in V}, \mathbf{w}) \mapsto y_{\G}$, where $\mathbf{w}$ represents trainable parameters, and the output $y_{\G}$ can be at either node or graph level label. In particular, Message Passing Neural Networks (MPNNs)~\citep{gilmer2017neural} are among the most studied classes of GNNs for their outstanding success in performing representation learning tasks on graph-structured data. MPNNs compute node representations through multiple rounds of message passing, as described by:

\begin{equation}\label{eq:message-passing-scheme}
\mathbf{h}_{v}^{\textsf{new}} = {\mathsf{com}}(\mathbf{h}_{v}, \underset{u \in \mathcal{N}(v)}{{\mathsf{agg}}} \big ( {\mathsf{M}} \big (\mathbf{h}_{u}, \mathbf{h}_{v} \big ) \big ) ,
\end{equation}

where $\mathsf{agg}$ denotes an aggregation function invariant to node permutations, and $\mathsf{com}$ combines the node's current state with messages from its neighbors. In \Cref{eq:message-passing-scheme}, $\textsf{M}$ is the message function that disseminates information across the neighborhood in $\G$.\\

\textbf{Graph Attention Networks:} 
Graph Attention Networks (GATs)~\cite{velivckovic2018graph} use a masked self-attention mechanism to prioritize and weigh the relevance of neighbors' features for each node when exchanging information in the message-passing scheme. For a graph $\G = (\V, \E)$ with node features ${\mathbf{x}}_v \in \mathbb{R}^d$, the importance of the latent representations of node $v$ is quantified by a scoring function $s_{\mathbf{w}} : \mathbb{R}^d \times \mathbb{R}^d \rightarrow \mathbb{R}$. Specifically, $s_{\mathbf{w}}$ is a parametric function computed as: $s_{\mathbf{w}}({\mathbf{x}}_v, {\mathbf{x}}_u) = \sigma \left( \mathbf{a}^\top \big[ \mathbf{W} {\mathbf{x}}_v \parallel \mathbf{W} {\mathbf{x}}_u \big] \right)$, where $\mathbf{W}$ is a learnable weight matrix, $\mathbf{a}$ is a learnable vector of attention coefficients, $\sigma$ is a point-wise non-linear activation function, and $\parallel$ denotes concatenation.
While the original GAT scoring function~\cite{velivckovic2018graph} employs a static attention mechanism, this can limit model expressiveness. To address this, a dynamic masked self-attention mechanism can be used by modifying the scoring function~\cite{brody2021attentive}: $s_{\mathbf{w}}({\mathbf{x}}_v, {\mathbf{x}}_u) = \mathbf{a}^\top \, \sigma \left( \mathbf{W}[{\mathbf{x}}_v \parallel {\mathbf{x}}_u] \right)$. Regardless of the chosen scoring function, it is crucial to ensure that the score magnitudes do not disproportionately affect aggregation, which can lead to unstable or biased learning. This is typically managed by scaling the scores using the \emph{softmax} function across neighbors: $\alpha_{v,u} ={\text{softmax}_{u \in \mathcal{N}(v)}} \,(s_{\mathbf{w}} ({\mathbf{x}}_v, {\mathbf{x}}_u))$.

This process ensures that the {\em normalized attention coefficients} are consistent across different neighborhoods. Additionally, it offers a probabilistic interpretation of the scores, enhancing our understanding of the model's attention distribution across $\G$. The normalized attention coefficients are then used to derive the latent representation of features as: ${\mathbf{h}}_v = {\mathsf{agg}_{u \in \mathcal{N}(v)}} (\alpha_{v, u} \, \mathbf{W} \, {\mathbf{x}}_u)$. To further enhance the expressiveness of graph attention and mitigate instabilities, $H$ distinct attention operations can be employed. These independent process relationships within the upper and lower neighborhoods and aggregate results through concatenation, summation, or averaging: ${\mathbf{h}}_v = \mathsf{agg}_{ u \in \mathcal{N}(v)} (\mathsf{agg}_{h}\, (\, \alpha_{v, u}^h \, \mathbf{W}^h \, \, {\mathbf{h}}_u))$.\\

\textbf{Update and Readout:} Once the latent representations are acquired, they are integrated with the current features to form an updated representation: ${\mathbf{x}}_v^{\textsf{new}} = \mathsf{com} ( {\mathbf{h}}_v, {\mathbf{x}}_v)$. After applying graph attention over a suitable number of layers or rounds, denoted as $\textsf{L}$, the graph's representation is computed by aggregating the representations of all nodes: ${\mathbf{x}}_{G} = \mathsf{agg}_{v \in \V} \, (\, {\mathbf{x}}_v^{\textsf{L}} \, )$. The aggregation function $\mathsf{agg}$ can be a max, mean, or sum operation. This result is then passed to a dense layer to generate predictions.\\

\begin{figure}[t]
    \centering
    \includegraphics[width=\linewidth]{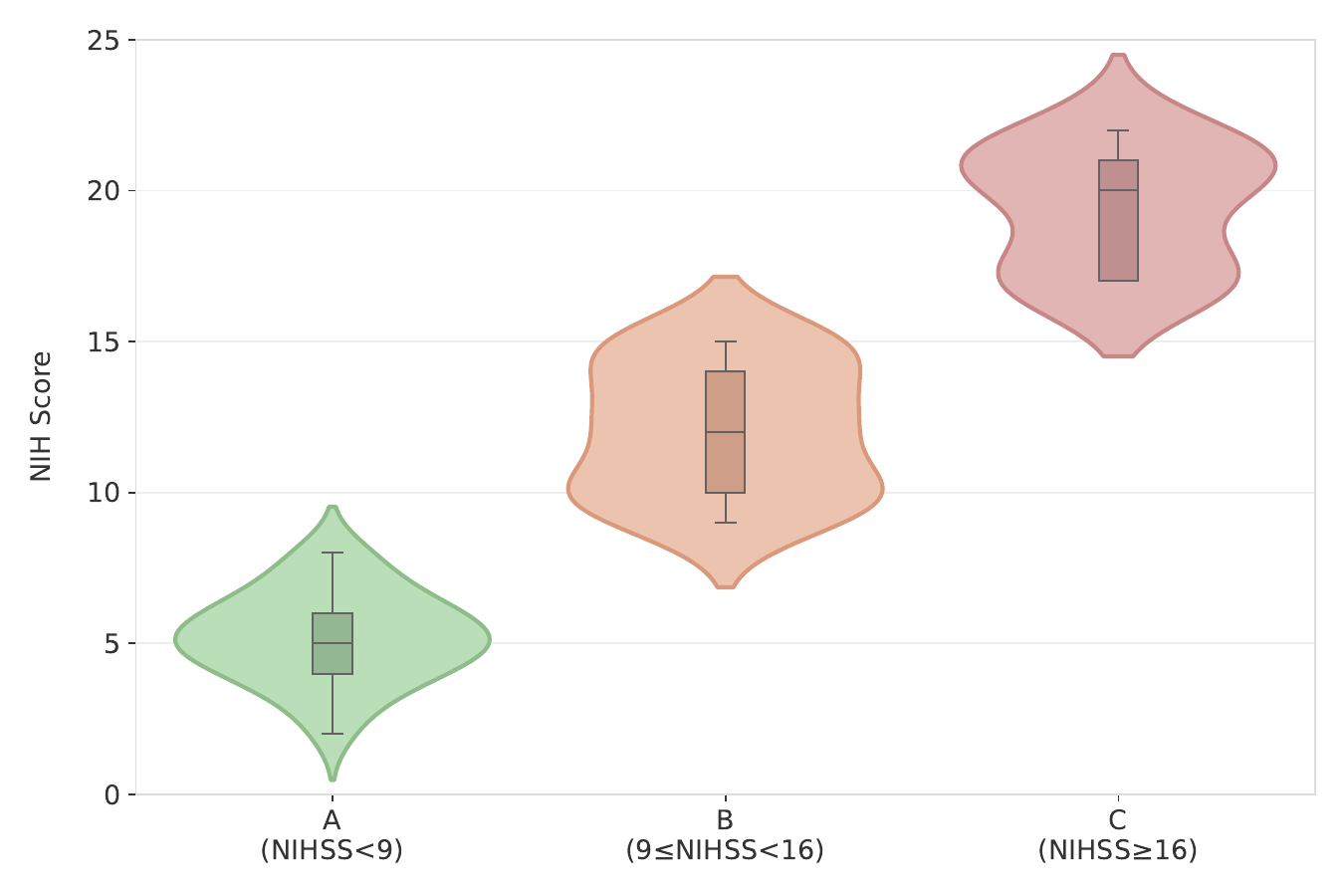}
     \caption{Distribution of NIHSS across three categorized severity groups: : Class A, for NIHSS < 9; Class B, for 9 $\leq$ NIHSS < 16; and Class C, for NIHSS $\geq$ 16} 
    \vspace{-3pt}
     \label{fig:nihh_distro}
\end{figure}

\textbf{Dataset Description:} Our study includes EEG recordings from 71 patients clinically assessed using the NIHSS at hospitalization. NIHSS scores in our dataset range from 2 to 22, distributed as illustrated in~\Cref{fig:nihh_distro}. For analysis convenience, we categorized patients into three classes: 

\begin{itemize}
    \item Class A: for all NIHSS < 9.
    \item Class B: for 9 $\leq$ NIHSS < 16.
    \item Class C: for NIHSS $\geq$ 16.
\end{itemize}

Note that this classification was created solely for our analysis and does not represent a standard method for categorizing stroke severity. EEG signals were recorded at rest, with eyes closed for at least 5 minutes during the stroke unit stay. Recordings were obtained from 31 electrodes positioned according to the international 10-10 system system with a common reference electrode placed on the mastoid and a ground electrode. EEG data were band-pass filtered from 0.2 to 47 Hz with a sampling rate of 512 Hz, and artifacts were removed using independent component analysis (ICA). The data were processed with eLORETA~\cite{pascual1994low} to reconstruct whole-brain sources and compute Lagged Linear Coherence (LLC) graphs for five frequency bands: $\delta$ (2-4 Hz), $\theta$ (4-8 Hz), $\alpha_1$ (8-10.5 Hz), $\alpha_2$ (10.5-13 Hz), and $\beta_1$ (13-20 Hz). In the absence of inherent node features, we used structural and positional encoding via Laplacian eigenvectors and random walk positional encoding. This method enabled us to input brain functional and structural connectivity into GNNs through brain graphs with node and edge attributes. \\

\textbf{Multi-layer Graph Attention Networks:} Inspired by the work in~\citep{casasroma2022applying}, which combines morphological, structural, and functional brain networks into a multi-layer graph representation, our experiments grouped LLC graphs of distinct frequency bands into a single multi-layer graph. This approach processes information from each band individually while allowing communication between nodes with the same label across different graph layers.

\begin{figure}[t]
  \centering
  \includegraphics[width=.7\linewidth]{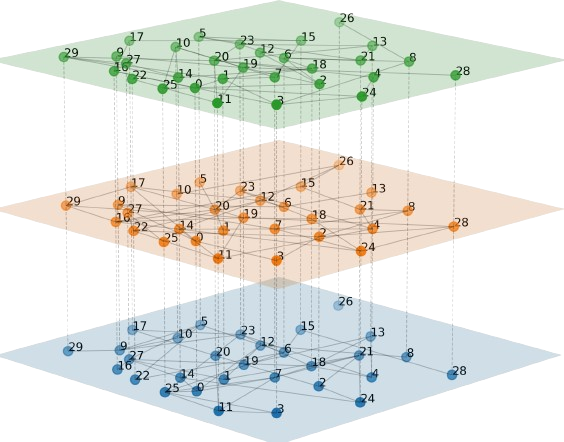}
  \caption{Illustration of a multi-layer network with three layers. Each layer is associated to a LLC graph for a specific frequency band. To allow for cross-layer communication, each node is connected to all the nodes with the same label of different layers. To reduce clutter, vertical dotted lines are placed only between successive layers.}
  \label{fig:multilayer_gnn}
\end{figure}

As illustrated in~\Cref{fig:multilayer_gnn}, the multi-layer graph, denoted as $\bar{\bar{\G}} = \{\G_1, \G_2, \ldots\}$, consists of multiple graphs. In our case, $\bar{\bar{\G}}$ comprises three layers corresponding to the $\alpha_1$, $\alpha_2$, and $\beta_1$ frequency bands. These bands were selected based on experiments testing all possible band combinations. The results indicated that using only the $\alpha_1$, $\alpha_2$, and $\beta_1$ bands provides the best configuration for noise reduction and improved NIHSS prediction accuracy. This choice aligns with the findings of~\cite{vecchio2019cortical}, which noted that high-frequency bands like $\alpha$ and $\beta$ often diminish after a stroke. This configuration allows targeted analysis of each band, represented by $\G_i = (\V_i, \E_i)$. Each node set $\V_i$ corresponds to the Brodmann areas, repeated across bands, tripling the total number of nodes in $\bar{\bar{\G}}$ to 252 nodes per patient.\\

\textbf{Rewiring Brain Networks:} Analysis of a single layer in our multilayer graph reveals a fully connected structure, which is suboptimal for GNN input due to potential over-smoothing issues \cite{rusch2023survey}. To address this, we developed a sparsification strategy.

Let $\G_l = (\V, \E_l)$ be a single layer of our multilayer graph $\bar{\G}$, where $l \in \mathcal{L}_{\text{final}}$. Our rewiring process transforms $\G_l$ into a sparse graph $\G_l' = (\V, \E_l')$, where $\E_l' \subset \E_l$, through two main steps:

\textit{Structural Rewiring:} we define a spatial proximity function $\phi: \V \times \V \rightarrow \mathbb{R}^+$ based on Euclidean distance between Brodmann areas. For each node $v \in \V$, we select the $k=3$ spatially closest nodes:
\begin{align}
    \mathcal{N}_k(v) = \{u \in \V \setminus \{v\} : |\{&w \in \V : \phi(v,w) \\ 
    &\leq \phi(v,u)\}| = k\} \nonumber
\end{align}
\begin{align}
    \E_{\phi} = \{ (u,v) \in \V \times \V : \ &u \in \mathcal{N}_k(v) \\
     \cup \ &v \in \mathcal{N}_k(u)\} \nonumber
\end{align} 

\textit{Functional Rewiring:} we define $\psi_l: \E_l \rightarrow \mathbb{R}$ mapping each edge to its LLC value in layer $l$. We retain edges above the $0.99^{th}$ quantile of LLC values:
    \begin{equation}
        \E_{\psi} = \{(u,v) \in \E_l : \psi_l(u,v) \geq Q_{0.99}(\psi_l)\}
    \end{equation}

The final set of edges in the rewired graph for layer $l$ is:
\begin{equation}
    \E_l' =  \E_{\phi} ~ \cup ~ \E_{\psi} \cup \{\underbrace{(v,v) : v \in \V\}}_{\text{self loops}}
\end{equation}

This process is applied independently to each layer $l \in \mathcal{L}_{\text{final}}$, resulting in a sparse multilayer graph $\bar{\G}' = (\V, \mathcal{L}_{\text{final}}, \bar{\E}')$.

Remarkably, after applying this sparsification process, on average only $\approx 5\%$ of the initial edges were retained. This significant reduction in edge density not only addresses the over-smoothing issue but also substantially reduces the computational complexity of subsequent GNN operations.

This rewiring strategy ensures that each layer maintains connections between proximal Brodmann areas and those with statistically significant LLC values, including self-loops. The resulting sparse graph structure enhances the GNN's ability to learn from the underlying brain network topology while preserving critical functional and structural information.

Figure \ref{fig:brain_network_plot} illustrates the full pipeline of a patient recording, in particular the described rewiring process, demonstrating the transformation from a fully connected graph to a sparse, rewired graph that combines structural and functional connections, retaining only $\approx 5\%$ of edges.

\section{Experiments}

\begin{figure*}[t]
    \vspace{-10pt}
    \centering
    \includegraphics[width=0.7\linewidth]{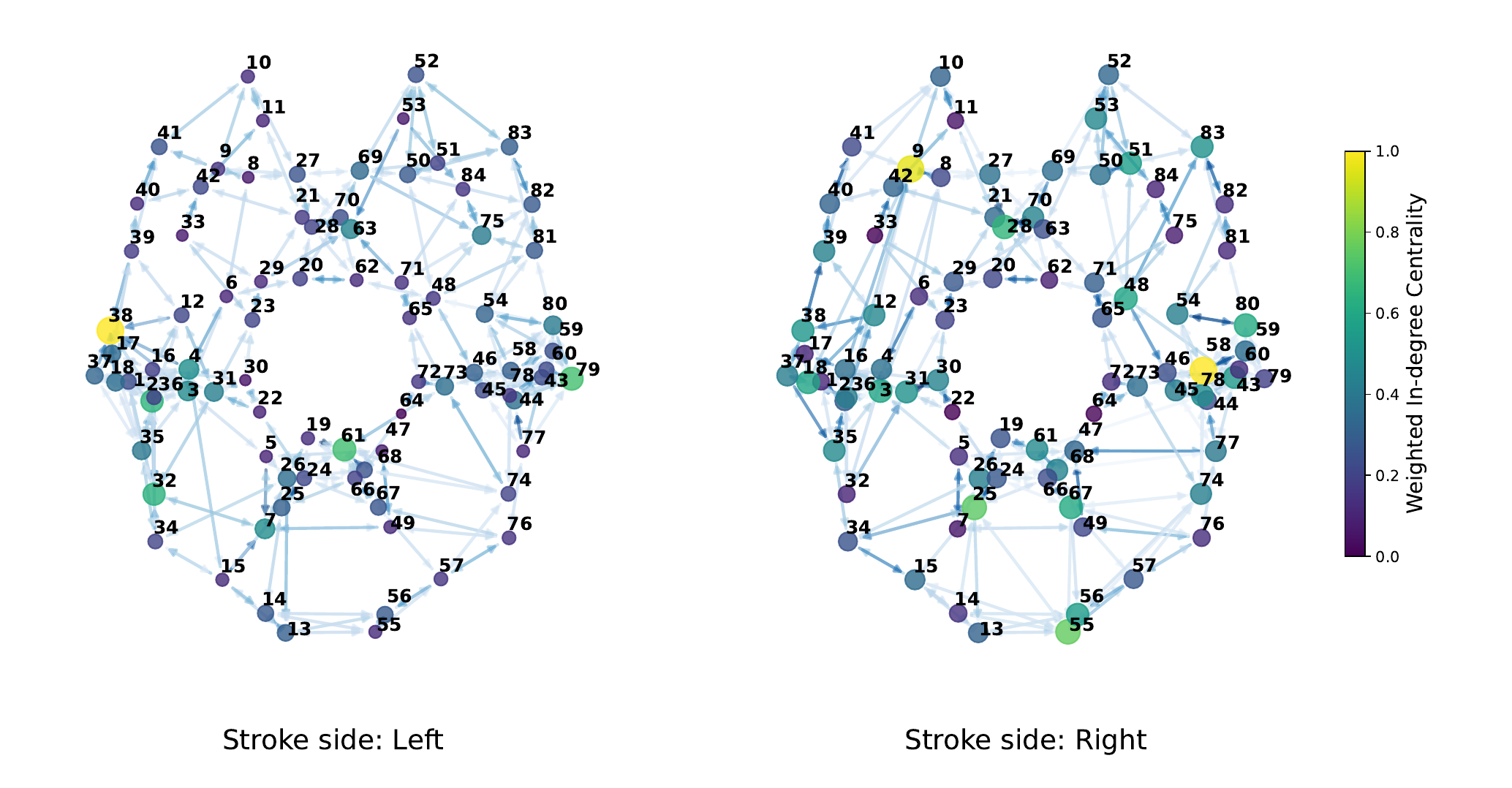}
     \caption{(Left) Patient A, Stroke side, left - NIHSS, 21. (Right) Patient B, Stroke side, right - NIHSS, 21. In both graphs, the color and size of the nodes are determined by the weighted in-degree centrality. These are the combined graph built from $G_{\alpha_1}$, $G_{\alpha_2}$, $G_{\beta_1}$. For each pair of Brodmann areas, the edge with the highest attention coefficient was selected: $e_{ij} = \max(e_{ij}^{\alpha_1}, e_{ij}^{\alpha_2}, e_{ij}^{\beta_1})$.}
    \vspace{-10pt}
     \label{fig:sx_vs_dx}
\end{figure*}

\begin{figure}[ht]
  \centering
  \vspace{-3pt}
  \includegraphics[width=\linewidth]{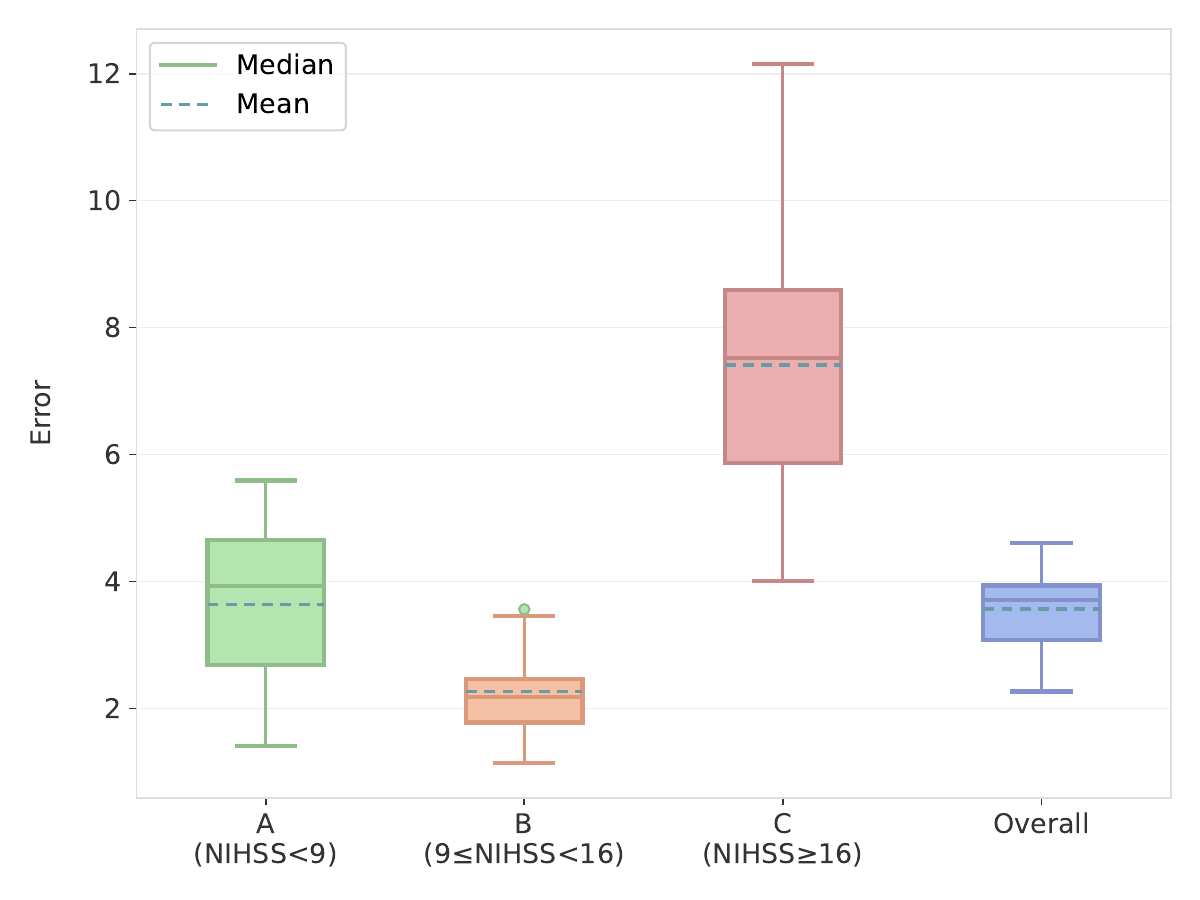}
  \caption{Error distribution overall and by class. Boxplots include a dashed line representing the mean of the distribution.}
  \label{fig:boxplot_mean}
\end{figure}

We tested a lightweight GATv2~\cite{brody2021attentive} model with only 62,593 parameters and a memory footprint of just 0.25 MB. The model consists of two layers, each with 64 hidden units and 8 attention heads, followed by ELU non-linearity. Despite its compact size, the model was trained efficiently with a batch size of 8 using the Adam optimizer. To speed up training and prevent overfitting, we incorporated Group Norm for normalization, applied a weight decay of $2.25\times10^{-4}$, and utilized Dropout with a 50\% probability of excluding units during training. The learning rate starts at $6.4\times10^{-3}$ and is halved whenever the validation loss plateaus, with patience of 20 epochs. For readout operations, we averaged the hidden features of each node across layers to obtain layer-wise representations, which were then concatenated and passed through a multi-layer perceptron for predictions. Our experiments employed k-fold cross-validation with five folds, reporting the Mean Absolute Error (MAE) at the point of early stopping. To ensure stable results, we repeated the process with five different random initializations of the model, yielding an average MAE of $\mathbf{3.57} \pm \mathbf{0.6}$ between the actual NIHSS and the predicted one.

In~\Cref{fig:boxplot_mean}, we perform a class-wise analysis of our model's results in terms of MAE of the NIHSS  across the three identified classes, along with the overall results. The results reflect our dataset distribution: the most accurate predictions correspond to class B ($\mathbf{2.2} \pm \mathbf{0.6}$), which is the most represented. In contrast, the scores of patients in class C ($\mathbf{7.4} \pm \mathbf{2.1}$) are often underestimated due to the low number of available observations, resulting in the highest error. For class A, we achieved results of $\mathbf{3.6} \pm \mathbf{1.2}$, which are slightly above the overall error due to a minor overestimation on average. These findings suggest that with a better dataset, specifically with a balanced distribution of NIHSS values, the error could significantly decrease below $3$, which is the estimated human error in defining the severity of a stroke event~\cite{schlegel2003utility}.

\begin{figure*}[t]
    \centering
    \includegraphics[width=\linewidth]{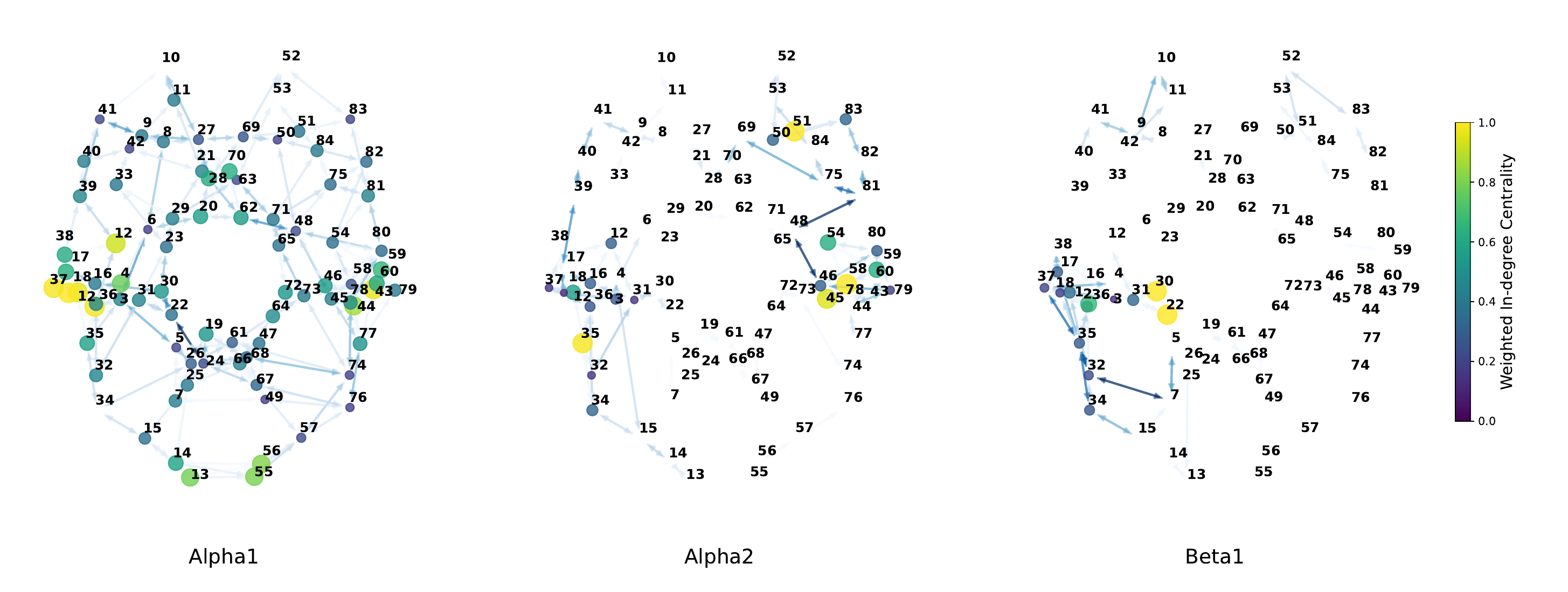}
     \caption{Using Patient A from \Cref{fig:sx_vs_dx} (left). Stroke side, left - NIHSS, 21. The weighted in-degree centrality determines the color and size of the nodes while the edges are based on the attention scores. This figure illustrates the utilization of the distinct frequency bands in the multi-layer graph.}
     \label{fig:bands}
\end{figure*}

\section{Related Works} 
The impact of acute stroke on the topology of cortical networks has been extensively investigated through EEG analysis, revealing significant, frequency-dependent alterations in network properties. Specifically, stroke leads to decreased small-worldness in the $\delta$ and $\theta$ bands and increased small-worldness in the $\alpha_2$ band across both hemispheres, regardless of lesion location~\cite{caliandro2017small}. 
Distinct modifications in functional cortical connectivity due to acute cerebellar and middle cerebral artery strokes have been highlighted, showing different impacts on network architecture and small-world characteristics across various EEG frequency bands, independent of ischemic lesion size~\cite{vecchio2019cortical}. Additionally, research has shown that acute cerebellar and middle cerebral artery strokes distinctly affect functional cortical connectivity, with significant differences in EEG-based network remodeling across $\delta$, $\beta_2$, and $\gamma$ frequency bands, highlighting the unique impact of stroke location on brain network dynamics~\cite{vecchio2019acute}. The prognostic role of hemispherical differences in brain network connectivity in acute stroke patients has been explored using EEG-based graph theory and coherence analysis. Findings indicate that stroke-induced alterations in network architecture can predict functional recovery outcomes, providing a basis for tailored rehabilitation strategies~\cite{vecchio2023prognostic}. In~\cite{guggisberg2019brain}, the relevance of brain network analysis for stroke rehabilitation has been studied, highlighting the potential of network-based approaches to inform and guide therapeutic interventions in stroke recovery. Dynamic functional reorganization of brain networks post-stroke has been emphasized, providing critical insights into the brain's adaptive mechanisms following a stroke and supporting the use of network analysis to understand structural and functional reorganization~\cite{wang2010dynamic}. Finally, changes in the contralesional hemisphere following stroke and the implications of the stroke connectome for cognitive and behavioral outcomes have been explored, enhancing our understanding of the complex network dynamics involved in stroke pathology and recovery~\cite{crofts2011network, lim2015stroke}.

\section{Conclusion}
This paper introduces a novel approach utilizing Graph Neural Networks (GNNs) to analyze, detect, and explain stroke severity based on EEG data collected from patients hospitalized in stroke units. The successful implementation of this method demonstrates the efficacy and robustness of graph representation learning for decision-making processes in stroke management. By leveraging attention coefficients embedded within the model, we not only accurately predict stroke severity but also provide insights into the reconfiguration process of the functional brain network. This provides clinicians with a valuable tool to support diagnosis and therapy evaluation.\\

\textbf{Interpretation of the results:} The analysis of the results obtained from our model demonstrates its effectiveness in predicting stroke severity based on EEG data. By examining the attention coefficients from the GNN, we gain valuable insights into the functional reorganization of brain networks post-stroke. Indeed, the model is able to highlight the most significant brain regions involved in the prediction process of clinical severity and therefore allows to deeply understand the relation between clinical findings and brain functional remodelling~(\Cref{fig:bands}). We need to consider that patients were not selected according to location and size of ischemic lesion in order to build a model as more generalizable as possible. Meanwhile, the diversity of extension and anatomical sites of ischemic lesions may be important elements in determining the value of attention coefficients. In this view, an important further implementation of our method will consist in integrating the functional multi-layer modelling with anatomical characterization of ischemic lesions and of their anatomical disconnection from other brain regions. This more comprehensive approach could allow a more precise prediction of clinical outcome and better depicting brain functioning. The goal of this work was not only to create a method that excels in predictions but also to ensure it is explainable. We achieved this objective through projection of the attention maps. In~\Cref{fig:g_theory_vs_attn}, we compare the previous approach based on small-world metrics with our method. It is evident that the use of GNN layers does not "destroy" the underlying connective structure but optimally leverages the connections by assigning them appropriate importance for NIHSS prediction. Lastly, it is interesting to compare two patients with strokes in opposite hemispheres. Stroke alters the connectivity in the affected regions, and \Cref{fig:sx_vs_dx} suggests that the model takes advantage of the changes of connections in these regions by assigning greater importance to these areas for NIHSS prediction. The figure clearly shows that the cluster of the most important Brodmann areas is on the right for the patient with a stroke in the right hemisphere and on the left for the patient with a stroke in the left hemisphere.\\

\textbf{Broader Impact:} The proposed use of Graph Neural Networks on brain graphs paves the way for addressing challenging tasks in clinical neuroscience, enabling prompt interventions and personalized therapies. This opens the path to having a more precise and timely assessment of stroke severity and identifying the brain regions that most significantly contributed to determining the patient's condition; our approach can help healthcare professionals tailor treatments and neuro-rehabilitation strategies more effectively, leading to improved patient outcomes and resource utilization.\\

\textbf{Future works:} We acknowledge that this work is limited by the dataset size, particularly for severe stroke cases. To address this, future research will focus on expanding the dataset and using federated learning techniques to facilitate this goal while addressing privacy concerns of healthcare institutions. In principle, we could extend our model's capabilities across multiple stroke units by leveraging federated learning without compromising patient data privacy and confidentiality. This collaborative approach enhances the scalability of our method and promotes cooperation among healthcare institutions, paving the way for more comprehensive and impactful research and development in stroke management.\\

\textbf{Implications for medical practice:} During the acute phase of ischemic stroke, the clinical condition can change suddenly and dramatically. Therefore, admitting patients into highly specialized stroke units that provide continuous multi-parametric monitoring is essential. The vital parameters usually detected in this context are blood pressure, oxygen saturation, and ECG, but no monitoring method allows for the fast detection of clinical changes. A quick detection of a clinical worsening could be decisive in the optimal management of patients. In this view, our approach based on EEG recordings and the ability to assess clinical severity could be a first step towards the functional and AI-mediated monitoring of stroke patients during their stay in the stroke unit. Moreover, formulating a precise prognosis for each stroke patient is a very complex task because a lot of variables can concur in conditioning brain functioning and clinical evolution over time. The GNN modeling of stroke-induced connectivity rearrangement could provide insight into short and long-term stroke prognosis and, therefore, contribute to tailoring a more personalized care path.

\bibliography{refs}

\end{document}